\documentclass[12pt]{article}
\usepackage{authblk}
\usepackage{amsmath}
\usepackage{float}

\usepackage{graphicx}
\usepackage{dcolumn}
\usepackage{bm}

\usepackage[natbib=true,style=numeric,sorting=none]{biblatex}
\usepackage[right=2.5cm, bottom=2.5cm, left=2.5cm, top=2cm]{geometry}

\usepackage{xcolor}

\usepackage{lineno}

\usepackage{amssymb}

\begin{document}

\title{Front End Amplifiers for Simultaneous Dual Readout of SiPM Sensors}

\author[1]{\small Lucas Finazzi\footnote{Corresponding author: lfinazzi@unsam.edu.ar}}
\author[1]{Felipe Soriano}
\author[1]{Leandro Gagliardi}
\author[1]{Alexis Luszczak}
\author[2]{Hugo Mériaux}
\author[1]{Gabriel Sanca}
\author[1]{Federico Golmar}

\affil[1]{Instituto de Ciencias Físicas, UNSAM-CONICET, Buenos Aires, Argentina}
\affil[2]{École nationale supérieure de l'électronique et de ses applications, Cergy Pontoise, France}

\date{\today}

\maketitle 

\begin{abstract}
    In this work, we present two fully open Analog Front End (AFE) designs for the simultaneous readout of the standard and fast outputs of ONSemi MicroFC-10035 Silicon Photomultipliers and report their main figure-of-merit parameters. The standard output AFE has a 1 p.e. amplitude of $(394 \pm 1)$~mV, which results in a Signal-to-Noise ratio of 30~dB. In addition, the fast AFE has a rise time of $(1.3\pm0.1)$~ns and a median electronic jitter of $34^{+7}_{-5}$~ps, resulting in an excellent timing performance. Using these amplifiers for simultaneous dual readout, we demonstrate a proof-of-concept application, in which fast output timing information is used to suppress correlated noise events or pileup events present in the standard output charge measurements. This suppression results in a reduction of non-prompt correlated noise (delayed crosstalks and afterpulses) in dark conditions and an increase up to $\sim$14 in single photon discrimination under a continuous incident photon rate of 7.8~MHz/mm$^2$. These results show that dual SiPM readout provides a practical way to remove correlated noise or pileup contamination in standard output charge measurements on an event-by-event basis.
\end{abstract}

\section{Introduction} \label{sec:intro}

Silicon photomultipliers (SiPMs) are optical detectors~\cite{sipm_review1, sipm_review2} that are capable of single photon detection. They are solid state, immune to magnetic fields, and operate at low voltages (typically between 25~V and 75~V, depending on SiPM brand). SiPMs are widely used in many diverse fields, including medicine~\cite{pet1, pet2}, particle physics~\cite{hep1, hep2, hep3}, quantum optics~\cite{qo1, finazzi_bunching_2024}, astrophysics~\cite{ap1, ap2}, space applications~\cite{gecam, finazzi_sipic_2026, finazzi_begonia_2024}, among others.

SiPMs are photon number resolving, so integrating an SiPM's signal over a fixed temporal window provides a charge/energy estimate of that event, which can then be translated to the total number of microcells fired. However, this measurement holds no timing information on when microcells fired within that integration window, so two events with the same total charge can have entirely different temporal structures and their difference could go unnoticed. Recovering this information requires access to the fast component of an SiPM signal, and some SiPM brands provide two separate outputs to address this issue: a standard output (typically 40~ns to 60~ns FWHM duration) with optimization for charge/energy resolution and a fast output (typically 1~ns FWHM duration), which is optimized to preserve the signal fast rising edge and minimize jitter for time-tagging. The energy linearity characteristic of the standard output and the timing resolution and high counting rate of the fast output are complimentary, rather than interchangeable. Dual readout of these two SiPM outputs could be used, for example, as an alternative to pulse-shape discrimination (PSD) using dual-gate charge integration of the standard output~\cite{psd1}. When using plastic scintillators, identification of different particles (gamma or neutrons, for example) relies on integrating the standard output with two gates of different width and then comparing the charge between them, as different particles result in a different time distribution of scintillation light. With dual readout, the standard output could be used to measure energy with only one integration gate, and the fast output could be used to identify different particle species depending on their measured time distribution.

In this work, we present two fully open Analog Front End (AFE) designs for the simultaneous readout of fast and standard outputs of ONSemi MicroFC-10035, and we perform a characterization of different figure-of-merit parameters. Secondly, we use this dual readout to demonstrate a proof-of-concept application, in which fast output information is used to suppress delayed optical crosstalks and afterpulses (when performing dark measurement) and pileup events (when illuminating the SiPM) to avoid event contamination in the standard output charge distribution. The paper is organized as follows: Section~\ref{sec:afe} describes the standard and fast AFE designs presented. Section~\ref{sec:setup} describes the experimental setup and the measurement methodology. Section~\ref{sec:results} presents the results obtained, and Section~\ref{sec:conclusions} presents the conclusions and comments on planned future work.

\subsection{Figures of Merit} \label{sec:fom}

There are two important figure-of-merit parameters used throughout this work. These quantify different aspects of the front-end electronics. The first figure of merit is the Signal-to-Noise ratio~\cite{snr}. This quantity is defined as 

\begin{equation}
    \mathrm{SNR} = \frac{A_{\text{1 p.e.}}}{\sigma_{\text{baseline}}} \, ,
    \label{eq:snr_std}
\end{equation}

\noindent where $A_{\text{1 p.e.}}$ is the 1 p.e. amplitude expressed in Volts and $\sigma_{\text{baseline}}$ is the pre-pulse baseline standard deviation (also in Volts). This SNR quantifies pulse amplitude in units of baseline noise and can be used to determine an acquisition optimal trigger level to avoid false detections and reduce pulse time walk.  

The second figure of merit, usually more important for fast output pulses, is the electronic jitter~\cite{jitter}. This jitter can be estimated as

\begin{equation}
    \sigma_t \simeq \frac{\sigma_{\text{noise}}}{\left. \text{dV}/\text{dt} \right|_{\text{level}}} \, ,
    \label{eq:jitter}
\end{equation}

\noindent where $\sigma_{\text{noise}}$ is the stationary electronic noise of the amplifier (usually the same as the baseline noise) and $\text{dV}/\text{dt} |_{\text{level}}$ is the pulse derivative for a given voltage level. This figure is the limiting factor in timing resolution achievable when using a front-end amplifier to time-tag an event, as voltage noise is converted into time jitter.

\section{Analog Front End} \label{sec:afe}

In this section we present both AFE designs used throughout this work to read out the standard and fast outputs of ONSemi MicroFC 10035 simultaneously.

\subsection{Standard Output} \label{sec:safe}

\begin{figure}[H]
    \centering
    \includegraphics[width=0.9\linewidth]{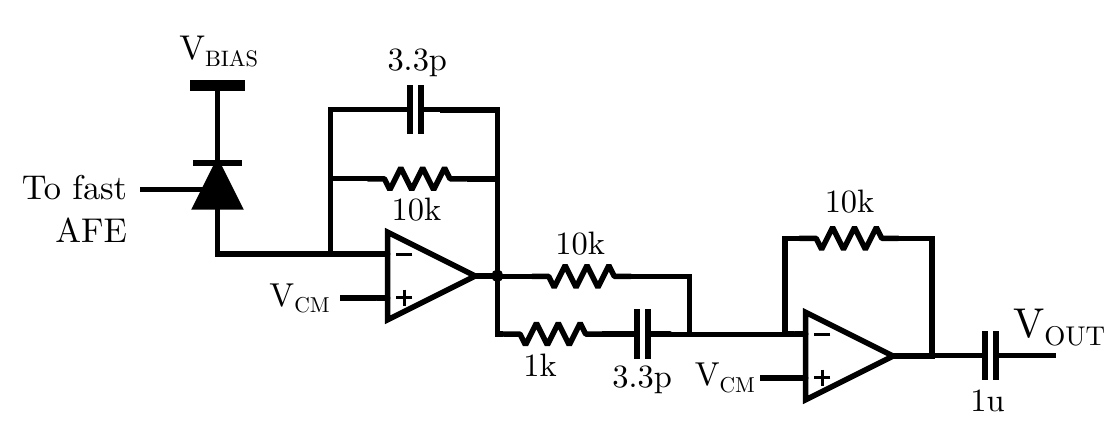}
    \caption{Analog Front End for the standard output of Silicon Photomultipliers. This design is comprised of a trans-impedance amplifier (TIA) as a first (inverting) stage and a voltage amplifier with an inverting gain of 21~dB as a second stage. The whole design has a common mode voltage of 4.5~V and a design bandwidth of 20~MHz. A 1~$\mu$F capacitor is used in the output to remove the common mode voltage component.}
    \label{fig:s_afe}
\end{figure}

This design was first presented in~\cite{muller_2020} and improved. Its current version is shown in Figure~\ref{fig:s_afe}.

The design is based in a trans-impedance amplifier (TIA) + voltage amplifier architecture using two OPA656 operational amplifiers biased with 9~V. The standard output of SiPMs consists of a current pulse that needs to be converted to voltage and amplified to be able to detect these events with ADCs or digitizers. The TIA stage is inverting and has a gain of 80~dB$\Omega$ and a 3.3~pF capacitor for feedback stability compensation. The second-stage voltage amplifier has a gain of 0~dB for frequencies $\lesssim 1$~MHz and a gain of 21~dB for frequencies above. This design amplifies pulse content while maintaining noise and low frequency components suppressed. The design bandwidth of this front end is 20~MHz. A common mode voltage of 4.5~V is used at the positive input of the operational amplifiers, which ensures that they operate in their linear region. It is important to note that the overvoltage of SiPMs read out by this amplifier is $V_{\text{over}} = V_{\text{bias}} - V_{\text{CM}}$. A 1~$\mu$F capacitor is used in the output to remove the common mode voltage component. The amplitude of a 1 p.e. pulse obtained with this design is approximately 400~mV. 

\subsection{Fast Output} \label{sec:fafe}

\begin{figure}[H]
    \centering
    \includegraphics[width=0.9\linewidth]{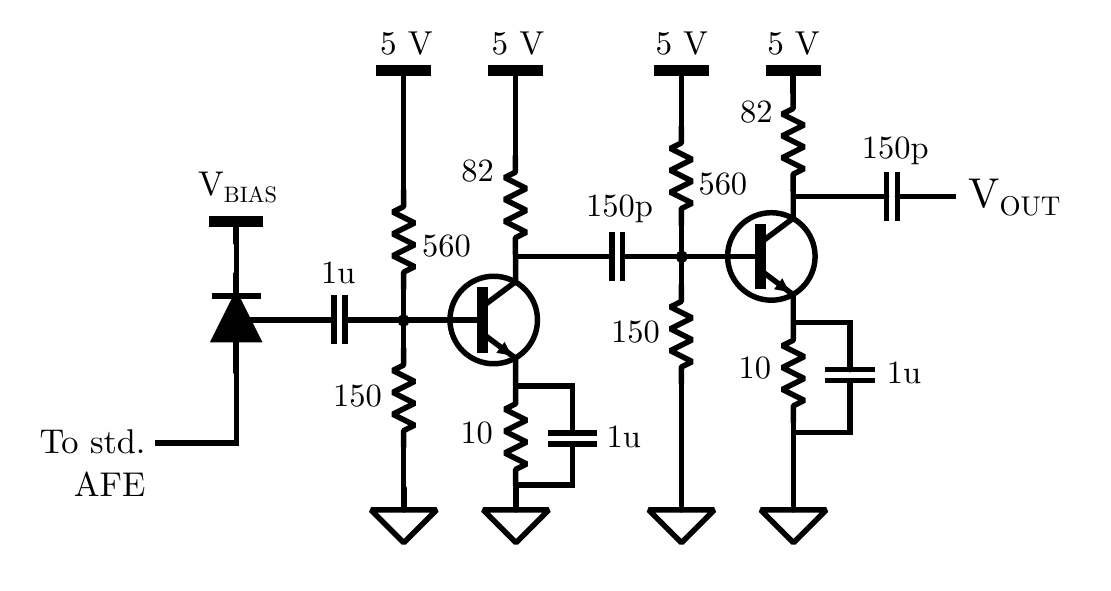}
    \caption{Analog Front End for the fast output of Silicon Photomultipliers. This design is comprised of two cascaded stages based on the BFU550XRR NPN transistor. The design gain for each individual stage is 20~dB. The transistors in this design are biased with 5~V through the base and collector resistors. A 10~$\Omega$ resistor was placed in the emitter to enhance transistor stability, with a parallel 1~$\mu$F capacitor to prevent further reduction in gain.}
    \label{fig:f_afe}
\end{figure}

The fast-output AFE improves on the design first presented in~\cite{soriano_2025}. Due to the high frequency content of fast pulses, transistors were chosen for this design instead of operational amplifiers. The design is based two cascaded voltage amplifier stages based on the BFU550XRR NPN transistor, as the fast output of Silicon Photomultipliers is a voltage output. The 2-stage design is shown in Figure~\ref{fig:f_afe}.

Each stage has an input and output impedance of approximately 50~$\Omega$, which allows them to be cascaded into several stages and mitigate ringing problems that can appear in high frequency amplifiers. Several transistors were evaluated for this design (Mini-Circuits MAR-6+, for example) but the BFU550XRR was ultimately chosen because the manufacturer provided spice models and S-parameters while other manufacturers did not. The transistors in this design are biased with 5~V through the base and collector resistors. A 10~$\Omega$ resistor was placed in the emitter to enhance transistor stability, with a parallel 1~$\mu$F capacitor to prevent further reduction in gain. The total design gain for this amplifier is 40~dB and the amplitude of a 1 p.e. pulse obtained with this design is approximately 70~mV. 

\subsection{Integration and Layout}

Both AFE designs were placed in the same PCB, which receives a power input of 9~V. A linear regulator was used to generate the 5~V needed for the fast AFE design. This prototype PCB had SMA connectors for signal input/output. A development board containing ONSemi MicroFC-10035 (MICROFC-SMA-10035-GEVB), which has SMA outputs for the standard and fast outputs, was used for AFE signal input through SMA cables. Some degraded performance was expected (as signals traverse transmission lines before measurement) when comparing with a board that has an on-board SiPM sensor close to the AFEs. To be able to use this development board with the standard AFE, the SiPM's anode 50~$\Omega$ resistor to ground was removed.

Four layers were used for the PCB, which allowed the signal-carrying traces of the fast AFE to be designed as coplanar waveguides~\cite{coplanar}, and to ensure that impedance stayed in the 50~$\Omega$ target throughout.

\section{Experimental Setup} \label{sec:setup}

For all measurements, the SiPM overvoltage used was 5~V and both standard and fast outputs were read simultaneously using the previously described Analog Front Ends. Measurements were performed at 24~$^\text{o}$C.

\subsection{Figure-of-merit Parameters}

To measure the different figure-of-merit parameters, a Tektronix MSO54 2~GHz oscilloscope was used. This is needed, especially for the electronic jitter measurement, as low bandwidth oscilloscopes or digitizers can modify the fast output pulses and degrade this parameter. A rising edge trigger at 15\% of the fast output 1 p.e. amplitude was used to trigger the oscilloscope. This value was low enough to avoid signal time-walk and high enough to avoid false triggers caused by the fluctuating baseline. A total of 1000 waveforms were acquired for both outputs and the following parameters were calculated after acquisition:

\begin{itemize}
    \item 1 p.e. amplitude,
    \item Derivative at 15\% of corresponding 1 p.e. amplitude.
    \item Baseline standard deviation (400 samples before primary pulse),
    \item Pulse FWHM,
    \item Pulse rise time (10\% - 90\%).
\end{itemize}

This way, a statistical distribution with 1000 entries was obtained for all these quantities and these were used to calculate the figure-of-merit parameters mentioned in Section~\ref{sec:fom}. For the electronic jitter figure of merit, no gaussian behavior was assumed and the reported values for this parameter are 68\% confidence intervals around the median value.

\subsection{Proof-of-concept Veto Procedure}

For this proof of concept, a CAEN DT5751 digitizer was used. Both outputs were acquired in coincidence mode with a coincidence window of 16~ns, as the average timing difference between both outputs is approximately 10~ns. A leading edge discriminator was used for both outputs using a level of 15\% of the output's corresponding 1 p.e. amplitude. The charge of the standard output was measured in a 70~ns integration gate (gain of 20~fC/ADC count) along with the fast output waveform in that same gate. This way, the charge/energy measurement in each event was paired with the fast output waveform, which contains temporal information on the generated charge. 

For each captured event, the number of peaks higher than the 15\% threshold was identified in each fast output waveform. Events with more than one fast output pulse were discarded because they correspond to either an event with an afterpulse, a delayed crosstalk, an accidental dark count, or a pile-up event. This veto procedure allows suppression of these events, which result in contamination of standard output charge. This procedure was performed in dark conditions (where most events discarded correspond to correlated noise) and for an incident photon rate of 7.8~MHz/mm$^2$ (where most events discarded correspond to pile-up events).

Detecting afterpulses can be difficult if they occur close to the main pulse ($\lesssim$ 15~ns). This is because their amplitude is small (as the microcell is recharging) and they can also get buried in the undershoot of the primary pulse. To be able to avoid this problem and to identify the afterpulses in dark measurements correctly, an undershoot correction algorithm was employed on the fast channel waveforms before the peak identification was performed on them. The correction algorithm consisted on:

\begin{enumerate}
    \item Calculating a template of the undershoot of pure 1 p.e., 2 p.e. (and higher) events, which was built with the average of 1000 waveforms without afterpulses or delayed crosstalks.
    \item Determining the amplitude of the primary fast pulse to correct to know which template to use in the algorithm.
    \item Subtracting the corresponding undershoot template from the raw waveform capture (starting 10~ns after the pulse maximum). The correction was performed this way to avoid amplitude reduction of the primary pulse.
\end{enumerate}

It is important to know that this undershoot correction procedure is needed in the proof-of-concept application in dark conditions, which targets events that have low amplitude by nature. However, this correction need not be necessary for other applications where the amplitude of events to cut or identify is of 1 p.e. or higher (for example, in the illuminated proof-of-concept application presented in this work).

To show that this undershoot correction didn't change the statistics of SiPM correlated noise, a 2D histogram of the timing separation between the primary and secondary events in a fast waveform vs. the secondary pulse amplitude was built. This histogram allows discrimination between different populations of SiPM correlated noise, like afterpulses and delayed crosstalks and several parameters can be read from it, like the microcell recharge time constant for the SiPM.

\section{Results} \label{sec:results}

\subsection{Figure-of-merit parameters}

\begin{figure}[H]
    \centering
    \includegraphics[width=0.8\linewidth]{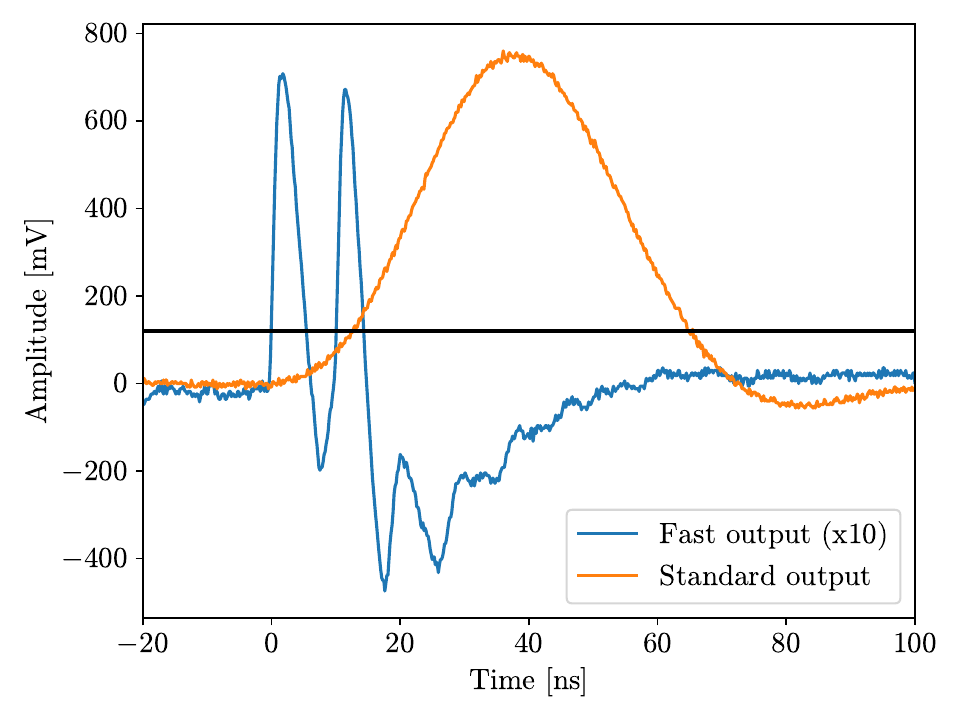}
    \caption{Example of fast and standard output waveforms for the same dark-count event. Two 1 p.e. events are discriminated by the fast output but are just seen as a 2 p.e. event on the standard output. The 15\% threshold (used as trigger level for measurements throughout this work) of the fast output is shown. Fast output signal is multiplied by 10 to visually compare both outputs.}
    \label{fig:waveforms}
\end{figure}

An oscilloscope capture of both SiPM channels for the same physical event are shown in Figure~\ref{fig:waveforms}.

It can be seen that two individual events can be separated by the fast output, while the standard output sees it as a 2 p.e. event (with no information if it was caused by direct crosstalk, delayed crosstalk, afterpulse or two genuine events). This Figure is representative of the 1000 oscilloscope captures performed to calculate the figure-of-merit parameters of the amplifiers. It can be seen that the fast output has an undershoot that can reduce the output of a pulse that falls after the primary. This effect is worse at a timing separation of $\sim$15~ns, but it is only relevant for low amplitude events, like afterpulses. Pulses that have 1 p.e. amplitude or higher are always detected even if they occur during the undershoot.

\begin{figure}[H]
    \centering
    \includegraphics[width=0.6\linewidth]{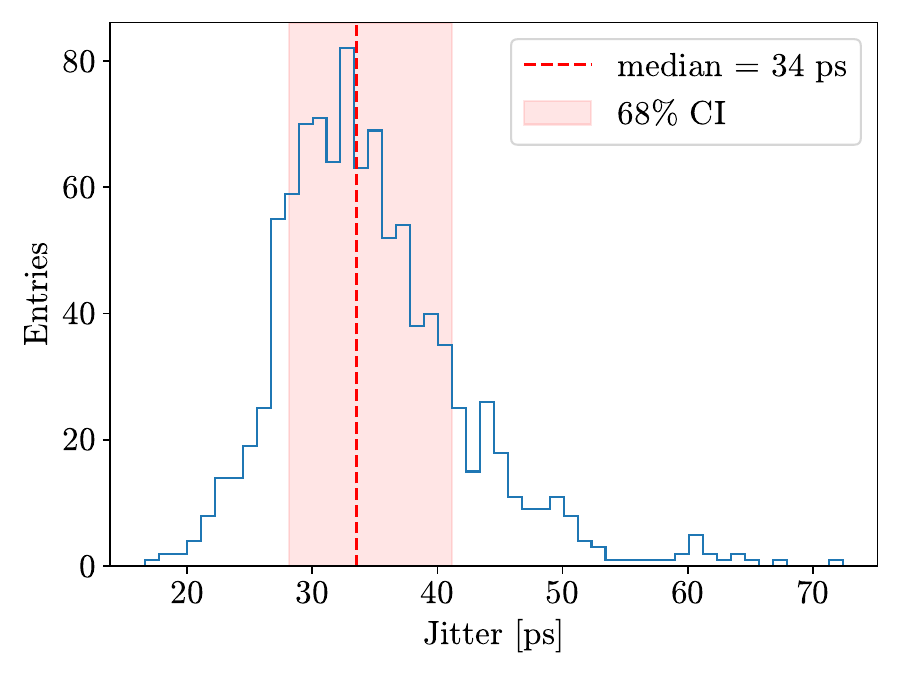}
    \caption{Histogram of fast output jitter using all pure 1 p.e. waveform captures (828 out of 1000) with the corresponding 68\% confidence interval marked. The median jitter is 34~ps, which results in an excellent single photon time resolution.}
    \label{fig:jitter_fom}
\end{figure}

\begin{figure}[H]
    \centering
    \includegraphics[width=0.55\linewidth]{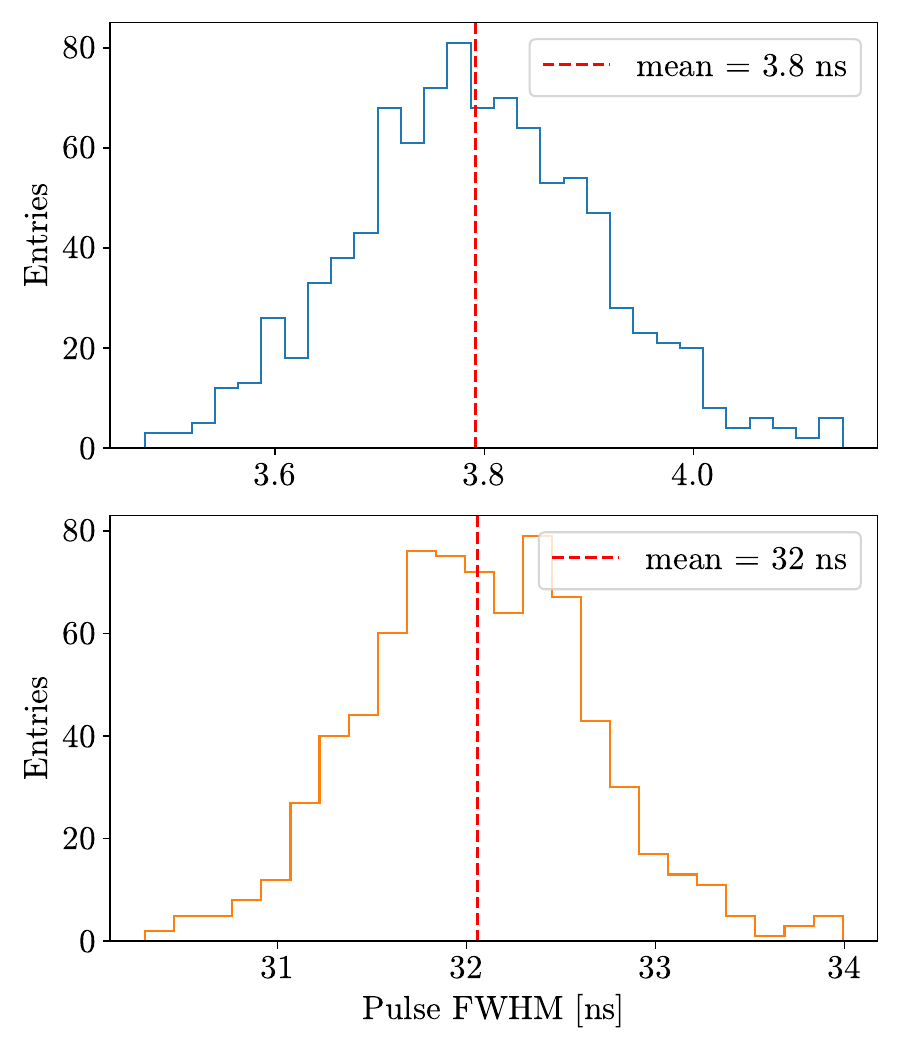}
    \caption{Histogram of pulse width for both outputs using all pure 1 p.e. waveform captures. The mean value for the fast output is 3.8~ns and the mean value for the standard output is 32~ns, which is 8.4 times larger. This value is reported in sigma due to the fact that the underlying distribution for this parameter is not known.}
    \label{fig:fwhm}
\end{figure}

Jitter measurement for the fast output and the pulse width for both outputs can be seen in Figure~\ref{fig:jitter_fom} and Figure~\ref{fig:fwhm}, respectively.

All the figure-of-merit parameters for both amplifiers are reported in Table~\ref{tab:afe_comparison}.

\begin{table}[H]
    \centering
    \caption{Comparison of the fast and standard AFE figure-of-merit parameters. All measurements were performed at 15\% of the corresponding 1 p.e. amplitude.}
    \label{tab:afe_comparison}
    \begin{tabular}{lcc}
        \hline
        Parameter & Fast AFE & Standard AFE \\
        \hline
        Pulse rise time (10\% - 90\%) [ns] & $(1.3 \pm 0.1)$ & $(18.6 \pm 0.1)$ \\
        Bandwidth [MHz] & 260 MHz & 19 MHz \\
        1~p.e.\ amplitude [mV] & $(70 \pm 1)$ & $(394 \pm 1)$ \\
        Pulse width (sigma) [ns] & $(3.8 \pm 0.1)$ & $(32 \pm 1)$ \\
        SNR [dB] & 33 & 30 \\
        Jitter (median, 68\% CI) [ps] & $34_{-5}^{+7}$ & $650_{-60}^{+50}$ \\
        Primary use & Timing/veto & Charge/energy \\
        \hline
    \end{tabular}
\end{table}

The main difference comes in the poor time resolution of the standard output when compared to that of the fast output. This is because the rise time of the fast output is more than 10 times faster than the standard output, which results in a larger derivative. In addition, the SNR for the fast AFE is larger due to the standard AFE having a noisier baseline. A 20~MHz low-pass filter could be used at the standard AFE output to remove a large part of this high frequency baseline noise and increase the SNR figure. It is important to note that performance will increase with an on-board SiPM, as this removes transmission lines (cables) from the experiment and reduces ringing and reflections, which reduce noise and timing performance.

\subsection{Proof-of-concept Veto Procedure}

\begin{figure}[H]
    \centering
    \includegraphics[width=0.9\linewidth]{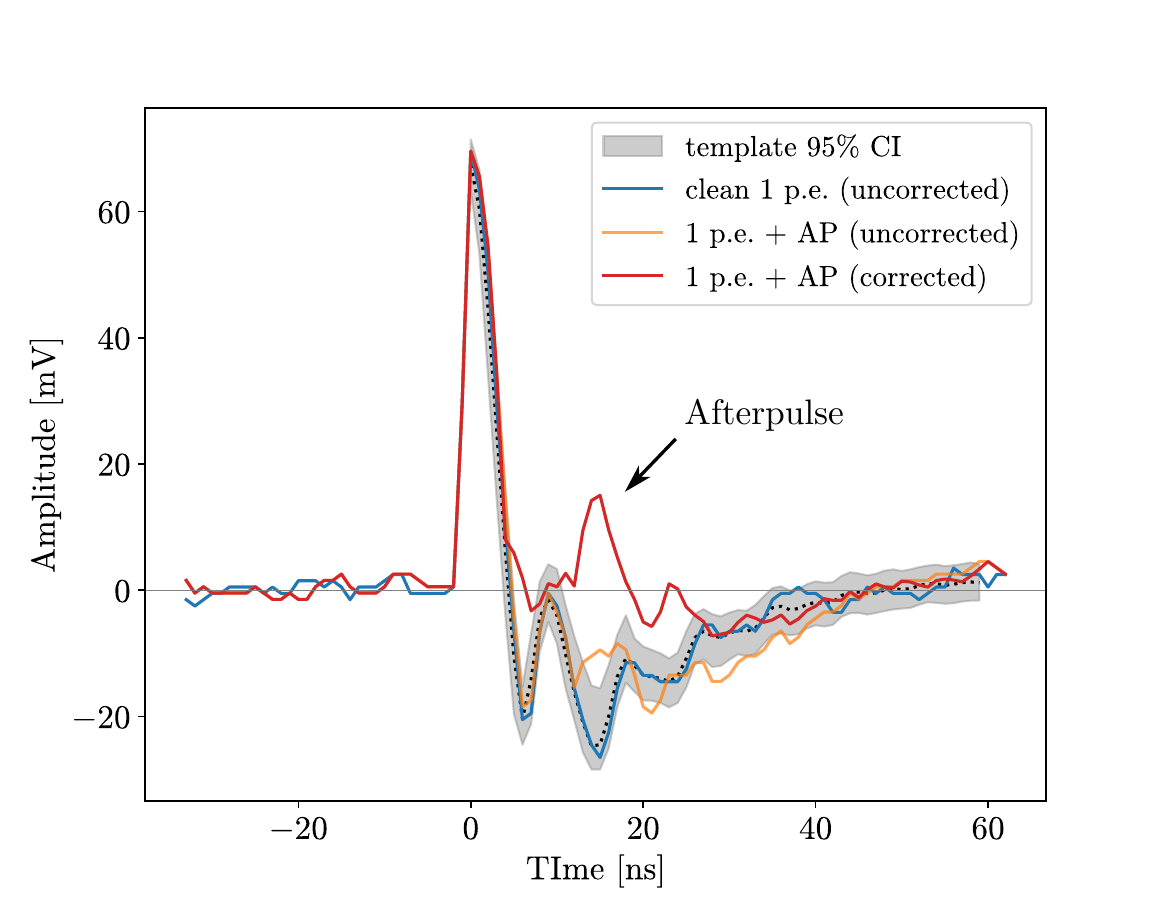}
    \caption{Template obtained from 1000 waveforms and the 95\% confidence interval along with many example pulses. A clean 1 p.e. is shown for reference, along with a clear/corrected 1 p.e. + AP event. It can be seen that afterpulsing events close to the primary can be as low as 20\% of 1 p.e. amplitude.}
    \label{fig:undershoot_corr}
\end{figure}

The comparison between an original waveform capture and a undershoot corrected signal is shown in Figure~\ref{fig:undershoot_corr}.

\begin{figure}[H]
    \centering
    \includegraphics[width=0.9\linewidth]{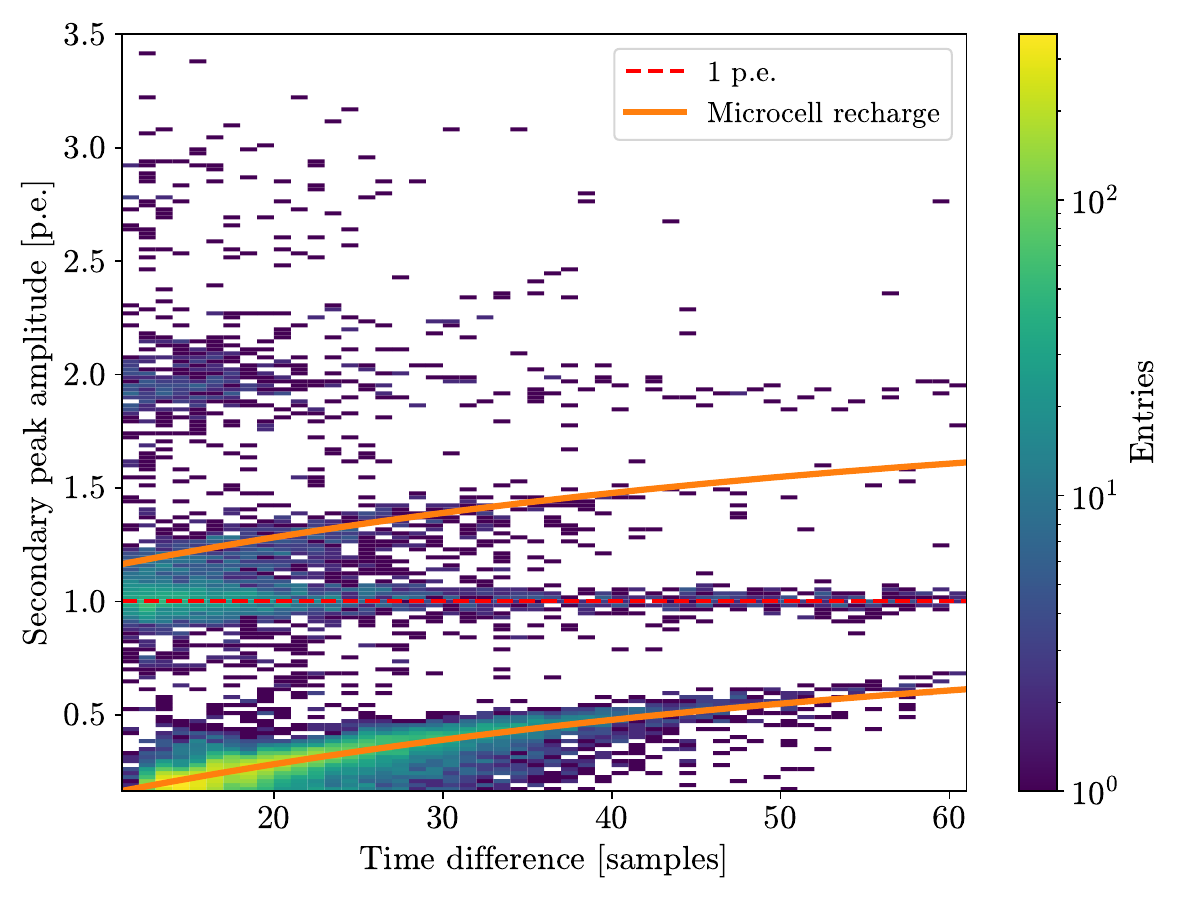}
    \caption{2D histogram of populations detected by the fast output. Delayed crosstalks can be seen as constant multiples of the 1 p.e. amplitude, while afterpulses clearly increase in amplitude for larger separation between the primary and secondary. An exponential fit was performed on events with amplitude lower than 0.6~p.e. (p-value of 0.68) to obtain the recharge time of a microcell. The value obtained by the fit was $(49 \pm 5)$~ns.}
    \label{fig:fast_pops}
\end{figure}

This figure shows how a corrected waveform makes detection of afterpulses possible. Afterpulse amplitude can be very small (< 20\% of 1 p.e. amplitude) for events separated $\sim$15~ns from the primary or less. To validate that this waveform correction procedure results in a proper identification of delayed SIPM populations, a 2D histogram of time difference between a primary and secondary fast pulse vs. the amplitude of the secondary pulse was built and it is shown in Figure~\ref{fig:fast_pops}.

In this Figure, two different populations can be observed. First we have delayed crosstalks, which are events with a constant amplitude and which don't depend on primary and secondary time separation. Secondly, we can observe afterpulses, which have an amplitude which is dependent on the time separation of fast events due to the influence of a recharging microcell. An exponential fit was performed on events lower than 0.6~p.e. (p-value of 0.68) to obtain the recharge time of a microcell. The obtained value was $(49 \pm 5)$~ns. The correct identification of these populations and the exponential recharge expected for afterpulses shows that the undershoot correction doesn't modify the statistics of the correlated noise and afterpulse events are possible after this correction.

\begin{figure}[H]
    \centering
    \includegraphics[width=0.9\linewidth]{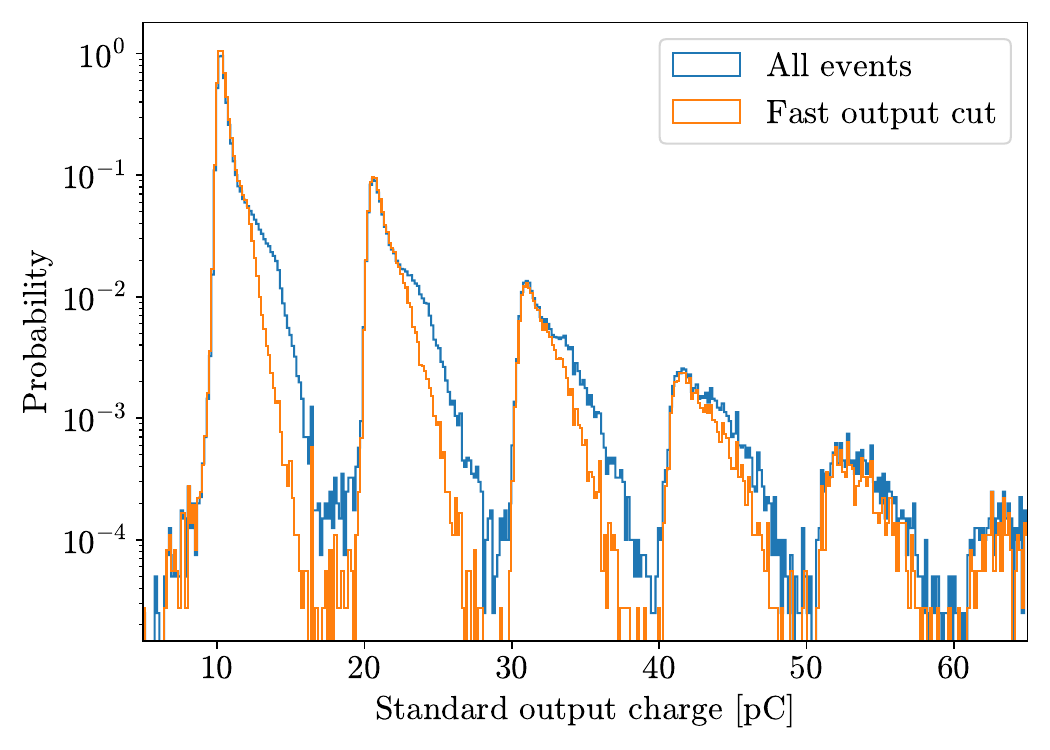}
    \caption{Finger spectrum in dark conditions of standard output charge before and after performing timing cuts with fast output information. It can be observed that a majority of right-shoulder events are removed naturally, as these are due to afterpulse contamination.}
    \label{fig:dark_cut}
\end{figure}

After the correction of fast output pulses, the veto procedure can be performed. The finger spectrum of the standard output charge in dark conditions (before and after the correlated noise cuts) is shown in Figure~\ref{fig:dark_cut}.

It can be observed that a majority of right-shoulder events are removed naturally from the finger spectrum, as these events are due to afterpulse contamination of the primary event. The probability at each p.e. peak is slightly modified as well, as this procedure cuts delayed crosstalk events as well. It is important to know that this method cuts a fraction of afterpulses, as very fast ones ($\lesssim$5~ns) are not detectable by this method due to these events being extremely low amplitude and overlapping with the primary pulse. The ratio of $\geq$2 p.e. to all events can be calculated from both histograms in Figure~\ref{fig:dark_cut} to assess the number of contaminated events that were cut. When all events are included, this ratio is 18.9\%, which corresponds roughly with the crosstalk probability for an SiPM~\cite{nepomuk_3_sipms}. After the cut was performed, this ratio was reduced to 16.5\%, indicating a flat reduction of 2.4\% (cut events are: all delayed crosstalks, afterpulses that occur more than $\sim$5~ns after the primary, and accidental dark counts). The probability to have two or more accidental dark counts in a 70~ns integration window at 24~$^\text{o}$C is 0.48\%.

\begin{figure}[H]
    \centering
    \includegraphics[width=0.9\linewidth]{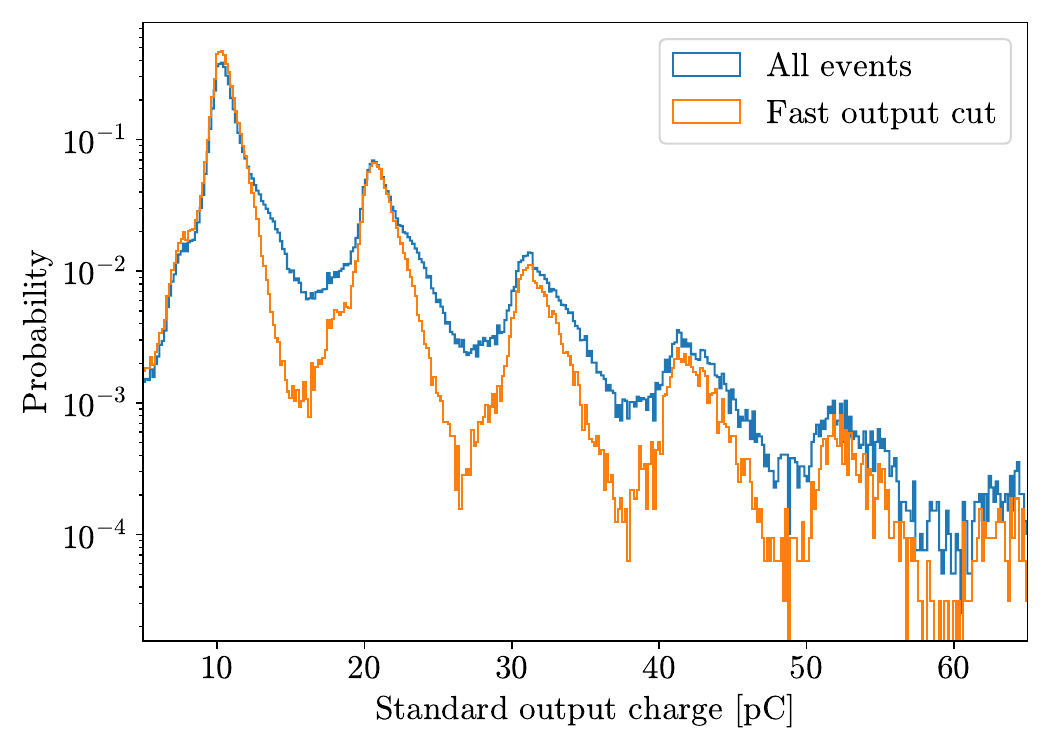}
    \caption{Finger spectrum of standard output before and after performing timing cuts with fast output information for an incident illumination of 7.8~MHz/mm$^2$. It can be observed that this cut suppresses events that fall between p.e. peaks (which are caused mainly by pileups, in addition to afterpulses and delayed crosstalks like in the dark case).}
    \label{fig:ill_cut}
\end{figure}

In addition, this proof-of-concept procedure was performed again with an incident continuous photon rate of 7.8~MHz/mm$^2$. This rate results in a non negligible pileup rate for the standard output (roughly 17\% for a 70~ns integration window), so a degradation of the single photon detection/performance of the sensor is expected for this illumination. This measurement is shown in Figure~\ref{fig:ill_cut}.

It can be shown that this cut suppresses events that fall between p.e. peaks (which are caused mainly by pileups, in addition to afterpulses and delayed crosstalks like in the dark case). This increases the ratio between local maxima and local minima of the finger spectrum, resulting in an increase in single photon resolution even at high illumination levels. Table~\ref{tab:pvr} shows the ratio between the p.e. peaks and the closest local minima with higher charge (before and after the cuts performed).

\begin{table}[htbp]
    \centering
    \caption{Ratio between the first three p.e. peaks and the closest local minima with higher charge (before and after the cuts performed). This value is often called Peak-to-Valley Ratio (PVR).}
    \label{tab:pvr}
    \begin{tabular}{c c c c}
    \hline
    Peak & PVR (before) & PVR (after) & Improvement (PVR after / PVR before) \\
    \hline
    1 & 63  & 620 & 9.8  \\
    2 & 31  & 427 & 13.8 \\
    3 & 19  & 184 & 9.7  \\
    \hline
    \end{tabular}
\end{table}

\section{Conclusions and Outlook} \label{sec:conclusions}

We have presented two full open Analog Front End designs for the simultaneous dual readout of ONSemi MicroFC-10035 along with their figure-of-merit parameters. The standard output amplifier has a 1 p.e. amplitude of $(394\pm1)$~mV, which results in an SNR of 30~dB, while the fast output amplifier achieves a rise time of $(1.3\pm0.1)$~ns and a median electronic jitter of $34^{+7}_{-5}$~ps. This excellent time resolution provided by this fast AFE is due to a steep derivative at 15\% of its 1 p.e. amplitude. 

Using these two amplifiers for simultaneous dual readout, we demonstrated a proof-of-concept application in which fast output information is not only used for time-tagging an event, but also to remove contamination from standard output charge measurements. In the dark case, this resulted in a reduction of the non-prompt correlated noise like afterpulses and delayed crosstalks. The fraction of $\geq$2~p.e.\ events in the standard output charge spectrum was reduced from 18.9\% to 16.5\%. This procedure was also evaluated at moderate lighting conditions (7.8~MHz/mm$^2$ incident), showing that it is able to naturally remove pile-up contamination in the standard output charge measurements. The Peak-to-Valley ratio, which determines the single photon discrimination capability of the detector, was improved up to a factor of $\sim$14.

Future work will focus on testing the amplifiers presented in this work on SiPMs of larger areas (for example, ONSemi MicroFC-30035), as higher area SiPMs are often used when coupling to scintillators to maximize the collection of scintillation light. This application will be done with on-board SiPMs for usage with plastic scintillators. 

\nolinenumbers

\section*{Acknowledgements}

The authors acknowledge financial support from \mbox{ANPCyT PICT 2017-0984} ``Componentes Electrónicos para Aplicaciones Satelitales (CEpAS)'', \mbox{PICT-2019-2019-02993} \mbox{``LabOSat:} desarrollo de un Instrumento detector de fotones individuales para aplicaciones espaciales'' and \mbox{UNSAM-ECyT} FP-001.

\printbibliography

\end{document}